\documentclass[12pt]{article}
\newcommand{\be}{\begin{equation}}
\newcommand{\ee}{\end{equation}}
\newcommand{\bi}[1]{\vspace{-3mm} \bibitem{#1}}
\usepackage{epsfig,amsmath,amssymb,graphics,graphicx}

\begin{document}

\begin{center}
{\it Journal of Physics: Condensed Matter 20 (2008) 4145212}
\vskip 5 mm

{\large \bf Fractional Equations of Curie-von Schweidler and Gauss Laws}
\vskip 5 mm

{\large \bf Vasily E. Tarasov} \\

\vskip 3mm

{\it Skobeltsyn Institute of Nuclear Physics, \\
Moscow State University, Moscow 119991, Russia } \\
{E-mail: tarasov@theory.sinp.msu.ru}
\end{center}

\begin{abstract}
The dielectric susceptibility of most materials 
follows a fractional power-law frequency dependence 
that is called the "universal" response.
We prove that in the time domain this dependence gives differential equations 
with derivatives and integrals of noninteger order.
We obtain equations that describe "universal" Curie-von Schweidler and 
Gauss laws for such dielectric materials.
These laws are presented by fractional differential equations 
such that the electromagnetic fields in the materials 
demonstrate "universal" fractional damping.
The suggested fractional equations are common (universal)
to a wide class of materials, regardless of the type 
of physical structure, chemical composition or 
of the nature of the polarization.
\end{abstract}

%03.50.De Classical electromagnetism, Maxwell equations 
%%%45.10.Hj Perturbation and fractional calculus methods
%41.20.-q Applied classical electromagnetism 

\newpage
%%%%%%%%%%%%%%%%%%%%%%%%%%%%%%%%%%%%%%%%%%%%%%%%%%%%%%%%%%%%%%%%%%%%%%%%%%
\section{Introduction} 

A growing number of dielectric relaxation data show that 
the classical Debye behavior \cite{W1,W2,W3} is hardly 
ever observed experimentally \cite{Jo1,Jo2,Jrev,Ram}. 
Instead it has been derived \cite{Jo1,Jo2,Jrev,Ram} that power laws 
are a common feature of the dielectric response of most materials
for wide frequency ranges. 
The fact that different dielectric spectra are described 
by the power laws is confirmed in many measurements \cite{Jo1,Jo2,Jrev} 
for a wide class of various substances.
The dielectric susceptibility of most materials follows, 
over extended frequency ranges, a fractional power-law 
frequency dependence, which is called the law of "universal" response 
\cite{Jo1,Jo2,Jrev}. 
This law is found both in dipolar materials beyond 
their loss-peak frequency, and in materials where the polarization 
arises from movements of either ionic or electronic hopping charge carriers.
These power-law responses are
most easily displayed in terms of the dielectric susceptibility
$\tilde \chi(\omega)=\chi^{\prime}(\omega)-i\chi^{\prime \prime}(\omega)$
as a function of frequency $\omega$. 
It has been found \cite{Jo4,Jo6} that 
the frequency dependence of the dielectric susceptibility
%%%$\tilde \chi(\omega)=\chi^{\prime}(\omega)-i\chi^{\prime \prime}(\omega)$
follows a common universal pattern for virtually all kinds of materials. 
The behavior
\be \label{W-3}
\chi^{\prime}(\omega) \sim \omega^{n-1} , \quad 
\chi^{\prime \prime}(\omega) \sim \omega^{n-1} , \quad 
(0 < n < 1, \quad \omega \gg \omega_p) ,
\ee
and
\be  \label{W-4}
\chi^{\prime}(0)-\chi^{\prime}(\omega) \sim \omega^{m} , \quad 
\chi^{\prime \prime}(\omega) \sim \omega^{m} , \quad 
(0 < m < 1, \quad \omega \ll \omega_p) ,
\ee
where $\chi^{\prime}(0)$ is the static polarization 
and $\omega_p$ the loss-peak frequency, 
is observed over many decades of frequency. 
Expressions (\ref{W-3}) and (\ref{W-4}) serve 
as the definition of the universal response behavior. 

Note that a consequence of the power laws is
that the ratio of the imaginary to the real component
of the susceptibility is independent of frequency. 
The frequency dependence given by equation (\ref{W-3}) 
implies that the real and
imaginary components of the complex susceptibility 
$\tilde \chi(\omega)=\chi^{\prime}(\omega)-i\chi^{\prime \prime}(\omega)$
obey at high frequencies the relation 
\be \label{W-5}
\frac{\chi^{\prime \prime}(\omega)}{\chi^{\prime }(\omega)} =
\cot \left( \frac{\pi n}{2} \right) , \quad  (\omega \gg \omega_p) .
\ee
The experimental behavior of equation (\ref{W-4}) 
leads to a similar frequency-independent rule 
for the low frequency polarization decrement:
\be \label{W-6}
\frac{\chi^{\prime \prime}(\omega)}{ \chi^{\prime }(0)- \chi^{\prime }(\omega) }
=\tan \left( \frac{\pi m}{2} \right) , \quad (\omega \ll \omega_p) .
\ee
This being a unique consequence of Kramers-Kronig relations 
and does not depend on any particular physical process.

%%%%%%%%%%%%%%%%%%%%%%%%%%%%%%%%%%%%%%%%%%%%%%%%%%%%%%%%%%%%%%%%%%%%%%%%%%
\section{Fractional equations for laws of universal response}

For the region $\omega \gg \omega_p$, 
the universal fractional power law (\ref{W-3})
can be presented in the form
\be \label{chi-1}
\tilde \chi(\omega)= \chi_{\alpha} \, (i \omega)^{-\alpha} , \quad (0<\alpha<1) 
\ee
with some positive constant $\chi_{\alpha}$ and $\alpha=1-n$. 
Using 
\[ (i\omega)^{\alpha}=|\omega|^{\alpha} \, 
\exp \{i \, \alpha \, \pi \, sgn(\omega)/2\}, \]
it is easy to derive relation (\ref{W-5}).
The polarization density can be written as
\be \label{Ptr}
{\bf P}(t,r)={\cal F}^{-1} \left( \tilde {\bf P}(\omega,r) \right)=
\varepsilon_0 {\cal F}^{-1} \left(\tilde \chi(\omega)  \tilde {\bf E}(\omega,r) \right) =
\varepsilon_0 \chi_{\alpha} \,
{\cal F}^{-1} \left( (i\omega)^{-\alpha} \tilde {\bf E}(\omega,r) \right) ,
\ee
where $\tilde {\bf P}(\omega,r)$ is 
a Fourier transform ${\cal F}$ of ${\bf P}(t,r)$. 
Equation (\ref{Ptr}) can be represented by integrals
of noninteger order $\alpha=1-n$.
The fractional Liouville integral \cite{SKM,KST} is defined by
\[ (I^{\alpha}_{+}f)(t)=\frac{1}{\Gamma(\alpha)} 
\int^{t}_{-\infty} \frac{f(t') dt'}{(t-t')^{1-\alpha}} . \]
The Fourier transform ${\cal F}$ of this integral is given 
(see Theorem 7.1 in \cite{SKM} and  Theorem 2.15 in \cite{KST})
by the relation
\[ ({\cal F} I^{\alpha}_{+}f)(\omega)=
\frac{1}{(i\omega)^{\alpha}} \, ({\cal F}f)(\omega) . \]
As a result, the fractional power law (\ref{chi-1}) gives
\be \label{Alpha}
{\bf P}(t,r)=\varepsilon_0 \chi_{\alpha} \, (I^{\alpha}_{+} {\bf E})(t,r) ,  
\quad (0<\alpha<1) . 
\ee

%%%%%%%%%%%%%%

For the region $\omega \ll \omega_p$, 
the universal fractional power law (\ref{W-4}) can be presented as
\be \label{chi-2}
\tilde \chi(\omega)=\tilde \chi(0)-
\chi_{\beta} \, (i \omega)^{\beta} ,  \quad (0<\beta<1) 
\ee
with some positive constants $\chi_{\beta}$, $\tilde \chi(0)$, and $\beta=m$. 
It is not hard to prove that equation (\ref{W-6}) is satisfied.
The law (\ref{chi-2}) can be presented by
the fractional Liouville derivative \cite{SKM,KST} that is denoted by $D^{\beta}_{+}$. 
The differential operator $D^{\beta}_{+}$ of order $\beta$ is defined by the equation
\be \label{+Da}
(D^{\beta}_{+}f)(t)=\frac{\partial^k}{\partial t^k}(I^{k-\beta}_{+}f)(t)
=\frac{1}{\Gamma(k-\beta)} \frac{\partial^k}{\partial t^k}
\int^{t}_{-\infty} \frac{f(t') dt'}{(t-t')^{\beta-k+1}} , 
\quad (k-1 < \beta <k) . \ee
The Fourier transforms ${\cal F}$ of this derivative
(see Theorem 7.1 in \cite{SKM} and  Theorem 2.15 in \cite{KST})
is given by
%%%If $0<Re(\alpha)<1$ and $f(t) \in L_1(\mathbb{R})$, 
%%%or $1\le p < 1/Re(\alpha)$ and $f(t) \in L_{p}(\mathbb{R})$, then
\[ ({\cal F} D^{\beta}_{+}f)(\omega)=(i\omega)^{\beta} ({\cal F}f)(\omega) . \]
As a result,
the fractional power law (\ref{chi-2}),  gives the polarization density 
\be \label{Pd2}
{\bf P}(t,r)= \varepsilon_0 \, 
{\cal F}^{-1} \left(\tilde \chi(\omega)  \tilde {\bf E}(\omega,r) \right) 
\ee
in the form
\be \label{Beta}
{\bf P}(t,r)= \varepsilon_0 \tilde \chi (0) \, {\bf E}(t,r) 
- \varepsilon_0 \chi_{\beta} \, (D^{\beta}_{+} {\bf E})(t,r) ,  
\quad (0<\beta<1) . \ee

Equations (\ref{Alpha}) and (\ref{Beta}) can be considered as the
universal response laws \cite{Jo1,Jo2,Jrev} for time-domain. 
These equations allows us to derive
fractional equations for electric and magnetic fields.

%%%%%%%%%%%%%%%%%%%%%%%%%%%%%%%%%%%%%%%%%%%%%%%%%%%%%%%%%%%%%%%%%%%%%%%%%%
\section{Fractional equations of the Curie-von Schweidler law}

Using (\ref{Alpha}) and (\ref{Beta}), the polarization current density
\be \label{Jpol}
{\bf J}_{pol}(t,r)=\frac{\partial {\bf P}(t,r)}{\partial t}
\ee
can be described by the fractional equations
\be \label{Alpha-J}
{\bf J}_{pol}(t,r)=\varepsilon_0 \chi_{\alpha} \, (D^{1-\alpha}_{+} {\bf E})(t,r) ,  
\quad (0<\alpha<1) ,
\ee
and
\be \label{Beta-J}
{\bf J}_{pol}(t,r)= \varepsilon_0 \tilde \chi (0)\, D^1_t {\bf E}(t,r) 
- \varepsilon_0 \chi_{\beta} \, (D^{1+\beta}_{+} {\bf E})(t,r) ,  
\quad (0<\beta<1) . \ee
For constant electric field ${\bf E}(t,r)$ equations (\ref{Alpha-J}) 
and (\ref{Beta-J}) show that the time dependence of the relaxation of 
the polarization current density (\ref{Jpol})
after the sudden removal of a polarizing field 
follows the power laws, which is widely observed in practice \cite{Jo1}
and is known as the Curie-von Schweidler law \cite{CvS1,CvS2}.
For the changeable field ${\bf E}(t,r)$, equations (\ref{Alpha-J}) and (\ref{Beta-J}) 
can be considered as a generalization of the well-known Curie-von Schweidler law.
Let us consider some examples of this generalization.

(1) Using (\ref{Alpha-J}) and (\ref{Beta-J}),  
we can derive the usual Curie-von Schweidler law. 
The most elementary of the applied field ${\bf E}(t,r)$ is the step function
\be 
{\bf E}(t,r)=
\begin{cases} 
0 , & t<a ,
\cr 
{\bf E}_0 (r) , & t>a .
\end{cases}
\ee
In this case, equations (\ref{Alpha-J}) and (\ref{Beta-J}) give
\[ {\bf J}_{pol}(t,r)=\varepsilon_0 \chi_{\alpha} \, 
{\bf E}_0 (r) ( _aD^{1-\alpha}_t 1)(t) = \]
\be
=\varepsilon_0 \chi_{\alpha} \, {\bf E}_0 (r) \, \frac{(t-a)^{\alpha-1}}{\Gamma(\alpha)},   
\quad ( t>a, \quad 0<\alpha<1) ,
\ee
and 
\[
{\bf J}_{pol}(t,r)= 
- \varepsilon_0 \chi_{\beta} \, {\bf E}_0 (r) (_aD^{1+\beta}_{t} 1)(t)=
- \varepsilon_0 \chi_{\beta} \, {\bf E}_0 (r) 
\frac{(t-a)^{-\beta-1}}{\Gamma(-\beta)}, \quad (t>a, \quad 0<\beta<1) , \]
where we use \cite{SKM} the relation
\[ ( _aD^{\alpha}_t 1)(t)= \frac{(t-a)^{-\alpha}}{\Gamma(1-\alpha)}, 
\quad (t>a, \quad \alpha>0) . \]
Here $ _aD^{\alpha}_{t}$ is the fractional derivative
\be \label{aDa}
( _aD^{\alpha}_{t}u)(t)=\frac{\partial^k}{\partial t^k}( _aI^{k-\alpha}_{t}u)(t)
=\frac{1}{\Gamma(k-\alpha)} \frac{\partial^k}{\partial t^k}
\int^{t}_{a} \frac{u(t') dt'}{(t-t')^{\alpha-k+1}} ,
\ee
where $n=[Re(\alpha)]+1$, i.e., $n-1<\alpha< n$.

As a result, we obtain the usual Curie-von Schweidler law that is described by
\[ {\bf E}(t,r)= \varepsilon_0 \chi_{\alpha} \, 
{\bf E}_0 (r) \frac{(t-a)^{\alpha-1}}{\Gamma(\alpha)},   
\quad ( t>a, \quad 0<\alpha<1) ,  \]
\be
{\bf E}(t,r)= - \varepsilon_0 \chi_{\beta} \, {\bf E}_0 (r) 
\frac{(t-a)^{-\beta-1}}{\Gamma(-\beta)}, \quad (t>a, \quad 0<\beta<1) .
\ee

(2) The experimental applied field  ${\bf E}(t,r)$ can be presented as
\be \label{Esin}
{\bf E}(t,r)= {\bf E}_0 (r) \sin(\lambda t ) .
\ee
Using the relation \cite{SKM}: 
\[
D^{\alpha}_{+} \sin(\lambda t+\phi)=
\lambda^{\alpha} \, \sin(\lambda t+\phi+\alpha \pi/2), \quad (\alpha>0) ,
\]
equations (\ref{Alpha-J}), (\ref{Beta-J}) and (\ref{Esin}) give
\be \label{Alpha-J2}
{\bf J}_{pol}(t,r)=\varepsilon_0 \chi_{\alpha} \, {\bf E}_0 (r) 
\lambda^{\alpha} \, \sin(\lambda t+(1-\alpha) \pi/2) , \quad (0<\alpha<1) ,
\ee
and
\be \label{2Jpol}
{\bf J}_{pol}(t,r)=  \varepsilon_0 {\bf E}_0 (r) 
\Bigl[ \tilde \chi (0) \lambda \, \cos (\lambda t)  
- \chi_{\beta} \lambda^{1+\beta} \, \sin(\lambda t+(1+\beta) \pi/2) 
\Bigr] , \quad (0<\beta<1) .  \ee
Equation (\ref{2Jpol}) can be rewritten in the form
\be
{\bf J}_{pol}(t,r)=  \varepsilon_0 a(\beta) \, 
{\bf E}_0 (r) \sin \Bigl(\lambda t +b(\beta) \Bigr) ,
\ee
where $a(\beta)$ and $b(\beta)$ describe the amplitude 
and phase changes by the equations
\[ a(\beta)=\sqrt{A^2(\beta)+B^2(\beta)} , \quad 
b(\beta)= arctan \Bigl(\frac{A(\beta)}{B(\beta)}\Bigr) . \]
Here
\[ A(\beta)=\tilde \chi (0) \lambda - \chi_{\beta} \lambda^{1+\beta} \, \cos (\beta \pi/2) , \]
\[ B(\beta)= \chi_{\beta} \lambda^{1+\beta} \, \sin(\beta \pi/2) . \]

(3) For the applied field
\be 
{\bf E}(t,r)=
\begin{cases} 
0 , & t<a ,
\cr 
{\bf E}_0 (r) g(t) , & t>a 
\end{cases}
\ee
with some function $g(t)$, exact expressions for
the polarization current density ${\bf J}_{pol}(t,r)$ can be 
derived by using the list of fractional derivatives of the function $g(t)$
(see Tables 9.1-9.3 in \cite{SKM}).
For $g(t)=(t-a)^s$, where $s>-1$,
\be
_aD^{\alpha}_t g(t)=\,
_aD^{\alpha}_t (t-a)^s=
\frac{\Gamma(s+1)}{\Gamma(s+1-\alpha)} (t-a)^{s-\alpha} ,
\quad (s>-1, \ \alpha>0) .
\ee
The fractional derivative of $g(t)=\cos [\lambda (t-a)]$ is
\[
_aD^{\alpha}_{t} \cos \lambda(t-a) =\frac{(t-a)^{-\alpha}}{2\Gamma(1-\alpha)}
\Bigl[\ _1F_1(1,1-\alpha,i\lambda(t-a))+ \ _1F_1(1,1-\alpha,-i\lambda(t-a)) \Bigr] ,
\]
where $\ _1F_1(a,b,c)$ is a hypergeometric function \cite{Erdelyi}. 
For $g(t)=exp(-\lambda t)$, we use
\be
_aD^{\alpha} e^{-\lambda t}= 
e^{-\lambda t} (t-a)^{-\alpha} E_{1,1-\alpha}[-\lambda(t-a)] ,
\quad (0<\alpha<1) ,
\ee
where $E_{\alpha,\beta}[z]$ is the Mittag-Leffler function \cite{MLF}:
\be \label{MLf}
E_{\alpha,\beta}[z]=\sum^{\infty}_{k=0} \frac{z^k}{\Gamma(\alpha k+\beta)}.
\ee
If $\alpha=\beta=1$, then $E_{1,1}[z]=\exp (z)$,  
where $\Gamma(k+1)=k!$ for positive integer $k$.

As a result, fractional relations (\ref{Alpha-J}) and (\ref{Beta-J}) 
can be considered as a generalization of the formulation of 
the Curie-von Schweidler law from a constant electric field 
into changeable fields ${\bf E}(t,r)$.

%%%%%%%%%%%%%%%%%%%%%%%%%%%%%%%%%%%%%%%%%%%%%%%%%%%%%%%%%%%%%%%%%%%%%%%%%%
\section{Fractional Gauss's laws for electric field}

Time-domain laws are presented by the fractional 
integral and differential equations (\ref{Alpha}) and (\ref{Beta}).
Using the equation 
\be 
{\bf D}(t,r)=\varepsilon_0 {\bf E}(t,r)+{\bf P}(t,r) , 
\ee
and Gauss's law 
\[ div \, {\bf D}(t,r)=\rho(t,r) \] 
for the electric displacement field ${\bf D}(t,r)$, we get 
\be \label{EPr}
\varepsilon_0 div \, {\bf E}(t,r) + div \, {\bf P}(t,r)=\rho(t,r) .
\ee
Substitution of (\ref{Alpha}) and (\ref{Beta}) into (\ref{EPr}) gives
\be 
\varepsilon_0 Z(t,r)+\varepsilon_0 \chi_{\alpha} ( I^{\alpha}_{+}Z)(t,r)
=\rho(t,r), \quad (0<\alpha<1) ,
\ee
\be 
\varepsilon_0 \tilde \chi(0) Z(t,r)- 
\varepsilon_0 \chi_{\beta} ( D^{\beta}_{+}Z)(t,r)=\rho(t,r),  
\quad (0<\beta<1) ,
\ee
where $Z(t,r)=div{\bf E}(t,r)$. 
Using $( D^{\alpha}_{+}\, I^{\alpha}_{+}Z)(t,r)=Z(t,r)$ 
(see Lemma 2.20 in \cite{KST}), we obtain
\be \label{G1}
( D^{\alpha}_{+}Z)(t,r)+\chi_{\alpha} Z(t,r)=
\frac{1}{\varepsilon_0}( D^{\alpha}_{+}\rho)(t,r),  
\quad (0<\alpha<1) ,
\ee
\be \label{G2}
( D^{\beta}_{+}Z)(t,r)-
\frac{\tilde \chi(0)}{\chi_{\beta}} Z(t,r)=-
\frac{1}{\varepsilon_0 \chi_{\beta}} \rho(t,r),  
\quad (0<\beta<1) ,
\ee
which are fractional differential equations of Gauss's law 
for the electric field ${\bf E}(t,r)$.

For a fixed (stationary) region $R$ of medium,
we define the total electric charge
\be
Q(t)=\int_R \rho(t,r) dV .
\ee
The electric field ${\bf E}={\bf E}(t,r)$ passing through a surface $S=\partial R$ 
gives the electric flux
\be
\Phi_E(t) =\int_S ({\bf E},d{\bf S})=\int_R  div {\bf E} \, dV .
\ee
The integration of equations (\ref{G1}) and (\ref{G2}) 
over the region $R$ gives the fractional equations
\be \label{G3}
( D^{\alpha}_{+}\Phi_E)(t)+\chi_{\alpha} \Phi_E(t)=
\frac{1}{\varepsilon_0}( D^{\alpha}_{+}Q)(t),  
\quad (0<\alpha<1) ,
\ee
\be \label{G4}
( D^{\beta}_{+}\Phi_E)(t)- \frac{\tilde \chi(0)}{\chi_{\beta} } \Phi_E(t)=-
\frac{1}{\varepsilon_0 \chi_{\beta}} Q(t),
\quad (0<\beta<1) .
\ee
These equations represent the integral Gauss's laws
for the electric field in dielectric media. 
Note that $D^{\alpha}_{+}$ is the differential operator of order
$\alpha$ that is defined by equation (\ref{+Da}).
If ${\bf E}(t,r)$ is defined by
\be 
{\bf E}(t,r)=
\begin{cases} 
0 , & t<a ,
\cr 
{\bf E}(t,r) , & t>a ,
\end{cases}
\ee
then the operator $D^{\alpha}_{+}$ transforms into the operator 
$ _aD^{\alpha}_{t}$  that is defined by equation (\ref{aDa}). \\

Consider equations (\ref{G3}) and (\ref{G4}) in the form
\be \label{Barrett}
_aD^{\alpha}_t u(t)- \lambda u(t) =f(t), \quad 0< \alpha<1 ,
\ee
where $u(t)$ presents $Z(t,r)$ or $\Phi_E(t)$ 
%%%for $a \rightarrow - \infty$,
$\lambda$ is $-\chi_{\alpha}$ or $\tilde \chi(0) / \chi_{\beta}$,
the function $f(t)$ is $(1/\varepsilon_0) \ _aD^{\alpha}_{t}\rho(t,r)$ and 
$(-1/\varepsilon_0 \chi_{\beta}) \rho(t,r)$, or 
$(1/\varepsilon_0) \ _aD^{\alpha}_{t}Q(t)$ and
$(-1/\varepsilon_0 \chi_{\beta}) Q(t)$.
The fractional derivative $\ _aD^{\alpha}_{t}$ is defined by (\ref{aDa}). 
Note that
\[ (D^{\alpha}_{+}u)(t) =
\lim_{a \rightarrow -\infty} ( _aD^{\alpha}_{t} u)(t). \]

J.H. Barrett \cite{Barrett} in 1954 first considered 
the Cauchy type problem for a linear differential equation (\ref{Barrett})
with the initial conditions
\be \label{Pr2}
( _aD^{\alpha-1}_{t}u)(a)=C ,
\ee
where $\ _aD^{\alpha-1}_{t}=\ _aI^{1-\alpha}_{t}$ is 
the Riemann-Liouville fractional integral \cite{KST}. 
Note that any change in the past of the input function
of a fractional order system changes the future of the solution \cite{KS,FS,HP}.
Hence the past of such systems cannot be represented 
by a finite set of local conditions \cite{KS,FS,HP}, and the initial conditions
have the integral form (\ref{Pr2}).

Barrett proved in \cite{Barrett} (see also \cite{KST} Theorem 4.1 and Example 4.1.) 
that if $f(t)$ is an integrable function on $(a,b)$,
then the problem (\ref{Barrett}), (\ref{Pr2}) has the unique solution 
given by
\be \label{Sol1}
u(t)=C (t-a)^{\alpha-1} E_{\alpha,\alpha} [\lambda (t-a)^{\alpha}]+
\int^t_{a} (t-t')^{\alpha-1} E_{\alpha,\alpha}[\lambda (t-t')^{\alpha}] \, f(t') \, dt' ,
\ee
where $E_{\alpha,\alpha}[z]$ is the Mittag-Leffler function defined by (\ref{MLf}).
For $f(t)=0$, we obtain
\be \label{Sol2}
u(t)=C (t-a)^{\alpha-1} E_{\alpha,\alpha} [\lambda (t-a)^{\alpha}] . 
\ee
To consider the asymptotic behavior of the solutions (\ref{Sol1}) and  (\ref{Sol2}), 
we can use the integral representation \cite{Podlubny,GLL1,GLL2} 
of the Mittag-Leffler function:
\be \label{59}
E_{\alpha,\beta}[z]=\frac{1}{2 \pi i \alpha} 
\int_{\gamma(a,\delta)} \frac{e^{ \xi^{1/\alpha} } 
\xi^{(1-\beta)/\alpha}}{\xi -z} d\xi, \quad (1<\alpha <2) ,
\ee
where $\pi \alpha/2 < \delta < min\{\pi, \pi \alpha\}$.
The contour $\gamma(a,\delta)$ consists of two rays
$S_{-\delta}=\{\arg(\xi)=-\delta, |\xi|\ge a\}$ and  
$S_{+\delta}=\{\arg(\xi)=+\delta, |\xi|\ge a\}$,
and a circular arc 
$C_{\delta}=\{|\xi|=1, -\delta \le arc(\xi) \le \delta \}$.
Let us denote the region on the left from $\gamma(a,\delta)$
by $G^{-}(a,\delta)$. 
Then the asymptote of (\ref{59}) has the form \cite{GLL1,GLL2}:
\be \label{60}
E_{\alpha,\beta}[z]=-\sum^{\infty}_{k=1} 
\frac{z^{-k}}{\Gamma(\beta-\alpha k)}, \quad z \in G^{-}(a,\delta), 
\quad (|z| \rightarrow \infty),  
\ee 
and $\delta \le |\arg(z)|\le \pi$.
In our case, $z=\lambda(t-a)^{\alpha}$, $\arg(z)=\pi$.
As a result, we arrive at the asymptote of the solution, 
which exhibits power-like tails.
These power-like tails are the most important effect of 
the noninteger derivative in the fractional equations.

%%%%%%%%%%%%%%%%%%%%%%%%%%%%%%%%%%%%%%%%%%%%%%%%%%%%%%%%%%%%%%%%%%%%%%%%%%

\section{Conclusion}

In this paper, it has been shown that 
the electromagnetic fields in a wide class of 
dielectric materials must be described by 
differential equations of fractional order 
with respect to time.
The orders of these equations are defined by exponents 
of the "universal" response laws  
for frequency dependence of the dielectric susceptibility. 
A remarkable property of the dynamics described by the fractional
equations for electromagnetic fields is that the solutions have power-like tails.
The typical features of the "universal" electromagnetic phenomenon 
and the suggested fractional equations are common to a wide class of materials, 
regardless of the type of physical structure, chemical composition
or of the nature of the polarizing species, whether dipoles, electrons or ions.

For small fractionality $\alpha$ (or $\beta$), it is possible to use
a $\varepsilon$-expansion \cite{TZ2} over the small parameter. 
The suggested fractional differential equations,
which describe the fields in dielectric media with power-law
frequency-domain response, can be solved numerically.
There are several numerical methods 
to solve fractional equations 
(for example, see \cite{SKM,Numer1,Numer2,Numer3}).
In this paper we consider only general equations without numerical examples. 

The presented approach does not take into account the dielectric 
dispersion region around the loss-peak frequency $\omega_p$. 
This is a limitation of the fractional calculus formalism. 
Note that the universal dispersion is only valid over a finite frequency region.
There exist numerical methods without {\it a piori} assumptions 
to tackle frequency to time and time to frequency conversions using 
distribution of relaxation times approach. 
There are several papers on this topic by different authors 
(for example, see \cite{Ref1,Ref2,Ref3,Ref4,Ref5}).

Note that Tuncer has shown \cite{Ref6} that in 
binary mixtures Maxwell-Wagner polarization 
could lead to a universal dispersion on a finite frequency region. 
In that study the disorder in the system leads to 
a universal dispersion through the Gauss' law.

Note that it will be interesting to find a generalized physical explanation 
for fractional power laws. 
Unfortunately, after more than 30 years we still do not have an
explanation of the universal response from first principles. 
There are several papers on the fractional kinetic approach to the dielectric
universal response by different authors 
(for example, see \cite{Ref21,Ref22,Ref23}).

%%%%%%%%%%%%%%%%%%%%%%%%%%%%%

\end{document}